\documentstyle[epsfig,10pt,times]{cernart}
\begin{document} 
\renewcommand{\mathrm}[1]{\mbox{#1}}
\renewcommand{\textfraction}{ 0.05}
\renewcommand{\bottomfraction}{ 0.90}
\renewcommand{\topfraction}{ 0.90}
\vskip-1mm
\date{ 25 July 1997 }
\mbox{ITEP-PH-5/97, CPPM/97-3, INP-MSU 97-18/469}
\vskip-1mm
\title{On the search for muonic photons \\
           in neutrino experiments\\
      }

\vskip-3mm
\author{     V.A.~Ilyin\thanks{ilyin@theory.npi.msu.su}~$^\ast$~, \ \
             L.B.~Okun\thanks{okun@vxitep.itep.ru}~$^\dagger$~, \ \ 
             A.N.~Rozanov\thanks{rozanov@cppm.in2p3.fr}$^\ddagger$~  \\ \\
     {\small \it 
       $^\ast$~Institute of Nuclear Physics, Moscow State University,
                        Moscow 119899, Russia}\\
    {\small \it 
       $^\dagger$~Institute of Theoretical and Experimental Physics,
                        Moscow 117218, Russia}\\
    {\small \it 
       $^\ddagger$~CPPM, IN2P3-CNRS, Marseille, France}
       }

\vspace*{-1mm}
\begin{abstract}\vspace*{-3mm}
\noindent
Conserved muonic number may turn out to be a conserved muonic charge, coupled
to muonic photons, $\gamma_{\mu}$. Muons and muonic neutrinos would emit
$\gamma_{\mu}$'s, which might be discovered by analysing the
data from the past and future high energy neutrino experiments (like CHARM II,
CCFR, CHORUS, NOMAD, etc.). There are two sources of $\gamma_{\mu}$'s in these
experiments: ~1) internal bremsstrahlung in pion and kaon decays into $\mu$ and
$\nu_{\mu}$, which provide neutrino beams; ~2) external bremsstrahlung of muons
in the shielding of the neutrino beam. In both cases the $\gamma_{\mu}$'s
would pass freely through the shielding and produce narrow muonic pairs in the
neutrino detectors. These pairs could be distinguished from the trident events
$(\nu_{\mu}+ Z \to \nu_{\mu}+ \mu^++\mu^-+ Z )$
 by their kinematical properties:
small $p_t$ of the muon pair, small invariant mass, etc.
All the above processes are quantitatively analysed in this paper.

\end{abstract}

\section{Introduction \label{introduction}}

Speculations on the existence of massless vector particles (extra
photons) coupled to baryonic number \cite{1} and leptonic numbers
\cite{2}-\cite{7} have been under discussion for several decades. In
particular, three types of such photons -- electronic, $\gamma_e$,
muonic, $\gamma_{\mu}$, and  tauonic, $\gamma_{\tau}$, -- have been
considered recently in ref.~\cite{3,7} where the search for muonic
photons at high energy neutrino experiments was discussed. In the
present paper we undertake quantitative analysis of such a search by
considering both the emission and the detection of $\gamma_{\mu}$'s.

As it is well known, the beams of muonic neutrinos in high energy
neutrino experiments are formed in a long decay tunnel, where pions and kaons
decay
:
\begin{equation}
   \pi^+ \to \mu^++\nu_{\mu}\;,\;\; \pi^- \to \mu^-+\bar{\nu}_{\mu}
                                                    \label{1}
\end{equation}
\begin{equation}
   \mbox{K}^+ \to \mu^++\nu_{\mu}\;,\;\; \mbox{K}^- \to \mu^-+\bar{\nu}_{\mu}
                                                    \label{2}
\end{equation}
($\mbox{K}_{\mu 3}$ decays are less important).
The decay tunnel is followed by  a thick shielding of iron,
 earth and concrete,
which absorbs the beam of muons and all other secondary particles (hadrons and
photons). For instance, in the neutrino beam of the CERN SPS
the 185 meters of iron were followed by 173 m of earth, and again
 by 20m of iron
and 15m of concrete \cite{GBEAM}.
The beam of muonic  neutrinos passes through the shielding and produces
reactions in the neutrino detector: mainly inelastic scattering
 on nuclei due to charged
and neutral current interactions.

The idea of the search for muonic photons in neutrino experiments is based on
the fact that the coupling constant of the muonic photons $\alpha_{\mu}$ is
small compared to $\alpha = 1/137$:
\begin{equation}
           \alpha_{\mu}/\alpha < 10^{-5}\;.
                                               \label{3}
\end{equation}
This limit follows \cite{2,7} from the precision measurements of the
$(g-2)_{\mu}$ -- anomalous magnetic moment of the muon \cite{9}. Because of
smallness of $\alpha_{\mu}$, muonic photons would easily  penetrate the
shielding in the neutrino experiments and produce narrow muon pairs in the
detector due to the reaction of
 the muon pair
production by $\gamma_{\mu}$ in the Coulomb field of a nucleus with charge $Z$:
\begin{equation} 
\gamma_{\mu} + Z \to \mu^+ + \mu^-\; + Z. \label{4}
\end{equation} 
 On the other hand, the rates of emission of muonic photons and of
production of muon pairs by them would be also small, which does not promise an
easy hunt.

%


The main background for the process (\ref{4}) are the so-called tridents:  
\begin{equation} 
      \nu_{\mu} + Z \to \nu_{\mu} + \mu^+ + \mu^- + Z\;.
                                                    \label{5} 
\end{equation}

As we will show, the dimuon invariant mass distribution for
tridents is broader, and these dimuons have much larger $p_t$-imbalance than
in process (\ref{4}), which may help to cope with the trident background.
%
%

To get accurate numerical results for the processes discussed in this
paper we used the package CompHEP \cite{10} to obtain analytical
expressions for squared Feynman diagrams.  For these purposes a new
Lagrangian have been implemented to include new particle, muonic
photon, and its interaction with muon and muonic neutrino.
Then, for each process under the calculation, the FORTRAN code was
generated for squared matrix element using the corresponding CompHEP
option, as well as the phase space factor with proper regularization
of kinematical singularities (see \cite{CompHEPsing} for the
description of regularization).  The nuclear form-factors were
implemented in this code as additional factors multiplying the
matrix elements of reactions (\ref{4}) and (\ref{5}).
The convolution with the spectra of muonic photons,
muons or neutrinos was realized using the so called `structure
function' option of the CompHEP package.
The phase space integration was performed with Monte-Carlo integrator
BASES, while events were generated by the program SPRING, both programs
are combined in the package BASES/SPRING \cite{BASES}.  The statistical
error in our MC calculations was less than 1\%.
The results of CompHEP calculations are compared with the results of
special CERN Monte-Carlo program in section~\ref{monte-carlo} of this paper.

In this paper we present the theoretical basis for the search for muonic
photons in a generic high energy neutrino experiment. The actual analysis of
experimental data is possible only for collaborations, which run the
experiments.

\section{Emission of $\gamma_\mu$'s in pion and kaon decays. \label{intbrems}}

The calculation of the internal bremsstrahlung in process (\ref{1}) is
straightforward, but the resulting spectrum has a simple  analytical
form only if one neglects the kinetic energy of the muon compared to
$ (m_{\pi} - m_{\mu})$. Then for a pion decaying at rest we get:
\begin{eqnarray}
      dB(\pi \to \mu +\nu_{\mu}+\gamma_{\mu}) 
   &\equiv&
                \frac{d\Gamma(\pi \to \mu+\nu_{\mu}+\gamma_{\mu})}
                {\Gamma(\pi \to \mu\nu)} =              \nonumber \\
   &=&
     \frac{2\alpha_{\mu}}{\pi} \cdot \frac{dx}{x}\cdot
     \left[(1 - x + \frac{1}{2} x^2)\cdot\ln
     \frac{2(1-x)}{m_{\nu_\mu}/(m_{\pi}-m_{\mu})}-1
                   +x \right],
                                                      \label{7}
\end{eqnarray}
where $x=\omega/\omega_{max}$, $\omega$ is the energy of the muonic
photon and 
 $\omega_{max}=(m_{\pi}^2 -m_{\mu}^2)/2m_{\pi} \approx m_\pi - m_\mu$.
Expression (\ref{7}) is
infrared divergent, but this divergence has no adverse physical implications,
because muonic photons of vanishing energy cannot produce muonic pairs.
Another important feature of the expression (\ref{7}) is its logarithmic
dependence on the mass of the muonic neutrino. This logarithm is due to
configurations in which the photon and neutrino momenta are collinear. Thus, if
one is lucky, one can not only discover the muonic photon, but to measure the
mass of the muonic neutrino, even if it is extremely tiny and  otherwise
inaccessible.
\begin{table}[htb]
\begin{center}
\begin{tabular}{|c|l|c|c|c|}
\hline 
 Decay mode &\multicolumn{1}{|c|}{$x=\omega/\omega_{max}$} 
            & $m_{\nu_\mu}=100$ KeV & $m_{\nu_\mu}=1$ eV 
            & \raisebox{0ex}[3ex][3ex]{$m_{\nu_\mu}=10^{-10}$ eV }\\ 
\hline \hline
                             & 0.9   & 0.367 & 1.37  & 3.39\\
                             & 0.7   & 0.599 & 2.00  & 4.80\\
$\pi\to\mu\nu_\mu\gamma_\mu$ & 0.5   & 1.01  & 3.25  & 7.74\\
                             & 0.3   & 2.08  & 6.54  & 15.5\\
                             & 0.2   & 3.50  & 10.9  & 25.6\\
                             & 0.1   & 7.86  & 24.1  & 56.6\\
\hline
                             & 0.9   & 0.083 & 0.211 & 0.465\\
                             & 0.7   & 0.127 & 0.304 & 0.657\\
                             & 0.5   & 0.210 & 0.494 & 1.06 \\
                             & 0.4   & 0.289 & 0.675 & 1.45 \\
K$\to\mu\nu_\mu\gamma_\mu$   & 0.3   & 0.426 & 0.991 & 2.12 \\
                             & 0.2   & 0.712 & 1.64  & 3.51 \\
                             & 0.1   & 1.59  & 3.65  & 7.76 \\
                             & 0.07  & 2.35  & 5.37  & 11.4 \\
                             & 0.05  & 3.36  & 7.68  & 16.3 \\
                             & 0.04  & 4.25  & 9.70  & 20.6 \\
\hline
\end{tabular}
\end{center}
\caption{
Internal bremsstrahlung:  differential branching $dB/d\omega$ (in units of
$10^{-5} \mbox{GeV}^{-1}$, assuming $\alpha_\mu/\alpha=10^{-5}$) as a function of the
muonic photon energy, $\omega$, for three different values of the muon
neutrino mass, $m_{\nu_\mu}$, in the rest frame of decaying meson.
The maximum photon energy is
 $\omega_{max}=(m_{\pi/\mbox{K}}^2 -m_{\mu}^2)/2m_{\pi/\mbox{K}}$.
\label{tab:intbrems}}
\end{table}
Both above features (infrared and collinear) are characteristic
for the spectrum of internal bremsstrahlung in a general case, when the kinetic
energy of muon is taken into account. To obtain accurate numerical results for
the differential branching, in particular  in the case of K decay, we
evaluated analytically the squared matrix element with the help of the package
CompHEP \cite{10}.  Then, the integration over squared invariant mass of
neutrino and $\gamma_\mu$ was done analytically using the computer algebra
system REDUCE \cite{11}. The obtained expression was evaluated
numerically with
high accuracy by the corresponding REDUCE option. The results of these
 accurate numerical calculations,  differ from (\ref{7}) by a few percent
 in the  case of pion.
 For the pion and kaon decays they are presented in 
table~\ref{tab:intbrems} for a set of neutrino masses, assuming 
$\alpha_\mu=10^{-5}\alpha$.

A muonic photon with energy $\omega$ in the rest frame of a meson (pion or
kaon) and angle $\theta$ (in the same frame) with respect to the direction of
the meson beam has in the laboratory frame energy $E_{\gamma_\mu}$, where
\begin{equation} 
E_{\gamma_\mu} = \omega\gamma(1 + v \cos \theta)\;. \label{8}
\end{equation} 
Here $v$ is the velocity of the meson and $\gamma = 1/\sqrt{1-v^2}$.
 The spectrum of muonic photons is derived by integrating over
$d\cos\theta$ and convoluting with the spectra of decayed pions and kaons.
Photons are emitted  within a cone with an angle $\theta' \sim 1/\gamma$. Thus, one should take into account the angular acceptance of the
detector determined by the geometry of experiment (the length of the decay
tunnel and the position and cross dimension of the detector).
In this paper we use as an example the CHARM II neutrino detector
at the 450 GeV
CERN SPS neutrino beam. This detector is located at 882 m from the $\pi$/K
production target with the length of the decay tunnel of 414 m.
 The formation of
the $\gamma_\mu$ spectrum is discussed in  ~\cite{app:} in
details in the framework of a simplified approach to the simulation of the 
detector geometry and meson beam.
As an example, we used the meson spectra from ref.~\cite{GBEAM} with the effective
K/$\pi$-ratio equal 0.24.
\footnote{We made the calculations only for the neutrino component of 
the neutrino beam
(focused positive mesons). Similar calculations should be done for the
antineutrino admixture due to wrong sign mesons and for antineutrino beam
(negative mesons focused), where we expect similar results.
Also we note that in this semi-analytical approach we neglect angular
and radial distributions of mesons and assume the uniform distribution
of the meson decay point along the decay tube. The exact account for
 these effects and their correlations is possible by pure Monte-Carlo
simulations and is presented in section~\ref{monte-carlo}. 
}
%

\begin{figure*}\flushleft
\begin{minipage}[htb]{0.48\linewidth}
\mbox{ \epsfig{file=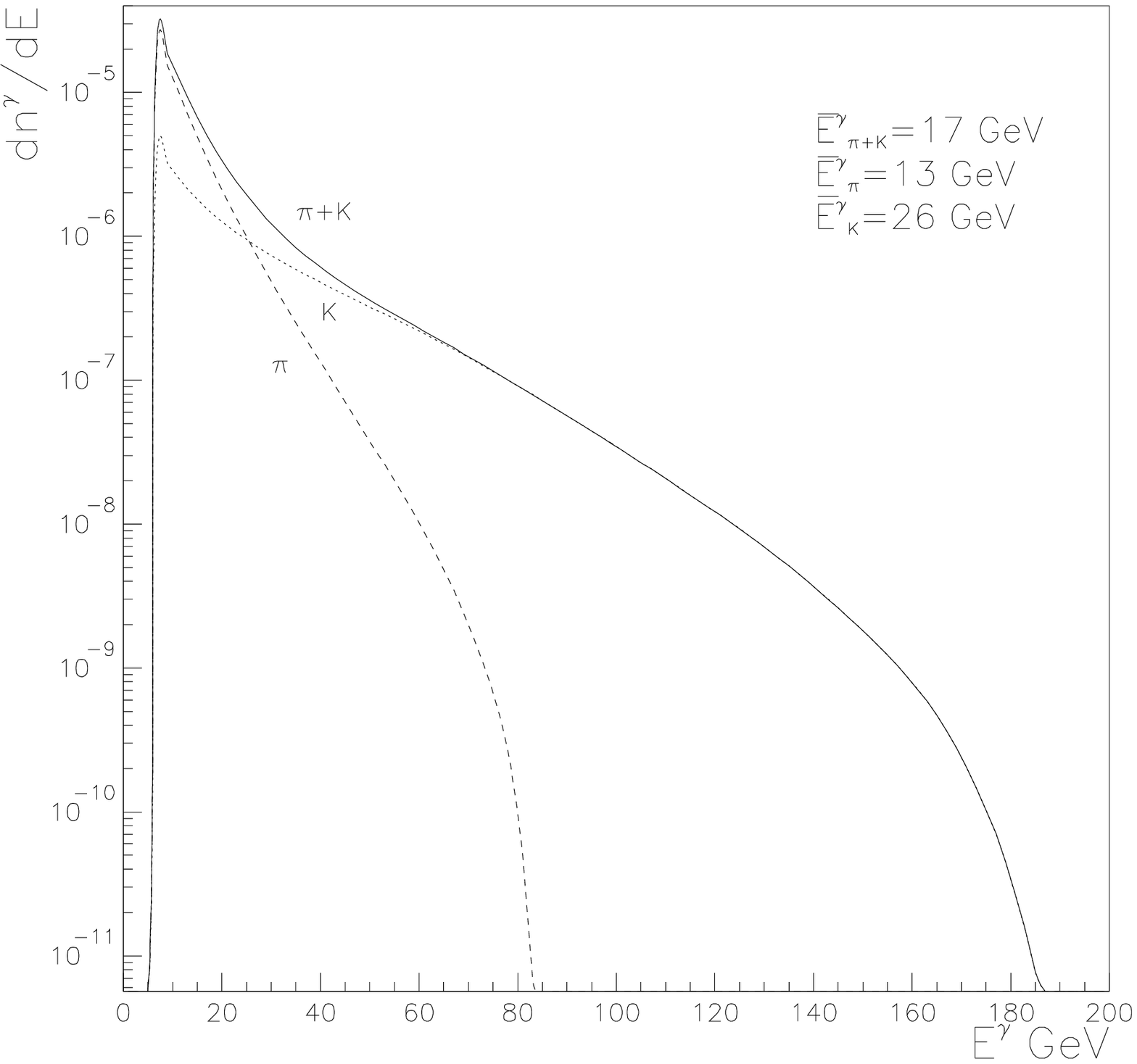,width=0.98\linewidth} }
\caption{
Spectra of muonic photons from internal bremsstrahlung in the CERN 450 GeV
  neutrino WBB (Wide Band Beam). The muon neutrino mass is assumed to be
 $m_{\nu}= 10^{-10}$ eV. }
\label{gint19sp}
\end{minipage}
\hspace{1mm}
\begin{minipage}[htb]{0.48\linewidth}
\mbox{ \epsfig{file=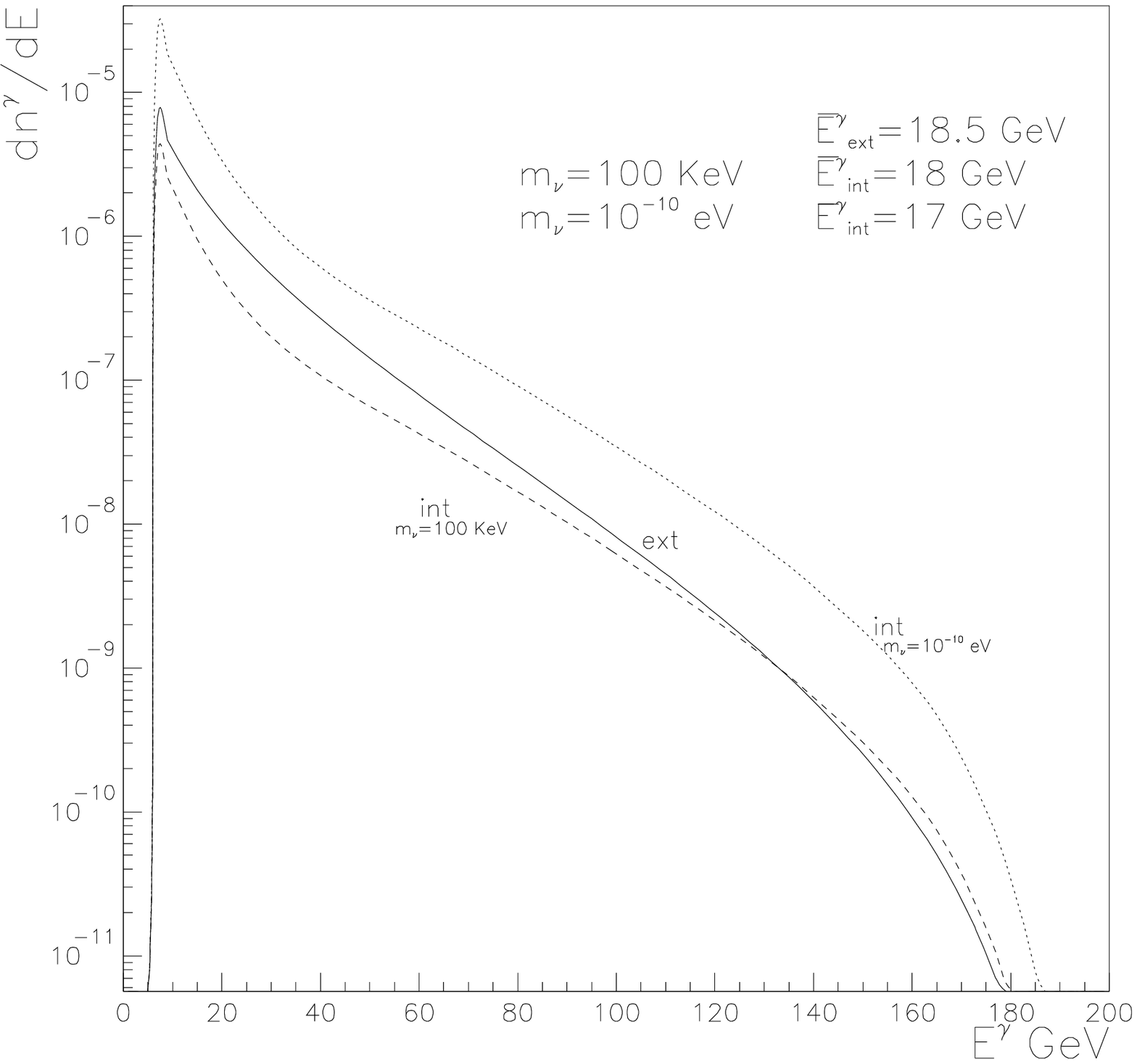,width=0.98\linewidth} }
\caption{
Comparison of the spectrum of muonic photons from internal bremsstrahlung
 in the CERN 450 GeV  WBB neutrino beam for two masses of muon neutrino
 (dotted line $m_{\nu}= 10^{-10}$ eV and dashed line $m_{\nu}= 100$ KeV ).
 The spectrum of muonic photons from external bremsstrahlung is shown by a
 solid line. }
\label{gallsp}
\end{minipage}
\end{figure*}

The resulting spectra of
muonic photons, produced by the internal bremsstrahlung, are presented in 
fig.~\ref{gint19sp}.

An important feature of fig.~\ref{gint19sp} is that hard  muonic
photons are mostly produced in the kaon decays. Of course, it depends on geometrical
parameters of the experiment (see ~\cite{app:}).  However this
feature could be taken into account  in future experiments to increase
possible  yield of muonic photons produced by the internal bremsstrahlung.

In order to see the dependence of rates and spectra on the mass of
 the muon neutrino we present in fig.~\ref{gallsp}
the spectra of muonic photons 
 for two values of the neutrino mass:  $ m_{\nu_{\mu}} = 100$
KeV - close to the
experimental limit \cite{9},  and another assumed to be very small,
$m_{\nu_\mu} = 10^{-10}$ eV.
The yield of muonic photons is about seven
times higher for the second choice of $m_{\nu_\mu}$. This allows one to
hope that the mass
of  muonic neutrino could be measured in the experiments discussed, if muonic
photon is discovered.

\section{Emission of $\gamma_\mu$'s by muons in the shielding. 
               \label{extbrems}}

The effective material for the muon bremsstrahlung
in the shielding is iron:
\begin{equation}
      \mu + \mbox{\rm Fe} \to \mu + \gamma_{\mu} + \mbox{\rm Fe}.
                                             \label{9}
\end{equation}

The cross-section of the bremsstrahlung for the electron has been calculated by
Bethe and Heitler \cite{12,13}; its derivation may be found in many text-books,
e.g. \cite{14,15}. In the case of muon there is an obvious substitution $m_e\to
m_{\mu}$. Moreover, the nuclear form-factor must be accounted for. This has
been done by a group at the Moscow Engineering Physics Institute (MEPhI)
\cite{16}-\cite{19}, see also \cite{20}-\cite{22}. In order to apply their
expressions to the process (\ref{9}) we have to multiply them by
$\alpha_{\mu}/\alpha$. In the ultra-relativistic limit, when $E_1\gg m_{\mu}$
and $E_2 \gg m_{\mu}$,  where $E_1(E_2)$ is the energy of the initial (final)
muon, we get:
\begin{equation}
     d\sigma \;=\; \alpha_{\mu}\frac{4}{3}
                   \left(\frac{2Z\alpha}{m_{\mu}}\right)^2 
                   \frac{dE_{\gamma_\mu}}{E_{\gamma_\mu}}
                   \left[1 - \frac{E_{\gamma_\mu}}{E_1} + \frac{3}{4}
                   \left(\frac{E_{\gamma_\mu}}{E_1}\right)^2\right] 
          \cdot \left[\ln \frac{2E_1(E_1-E_{\gamma_\mu})}
          {m_{\mu} E_{\gamma_\mu}}-\frac{1}{2} - 1.5 \right].
                                                     \label{10}
\end{equation}
Here $E_{\gamma_\mu}$ is muonic photon energy, 
$Z$ is the charge of a nucleus of iron $(Z = 26)$. The term 1.5 appears 
due to the nuclear form-factor, which was assumed in the 
form\footnote{An extensive compilation of nuclear elastic 
form-factors can be found in ref.~\cite{23}.}:
\begin{equation}
         F(q^2) = \exp(-q^2 a^2/6), \qquad
                   a = (0.58 + 0.82 A^{1/3}) \mbox{\rm fm}.
                                                     \label{11}
\end{equation}
Another para\-met\-ri\-za\-tion of the nuclear form-factor 
correction is suggested in ref.~\cite{19}, according to which, instead of 
1.5 in the second bracket of eq.~(\ref{10}), one has to substitute
\begin{equation}
    \Delta^{el}_n(\delta) =
            \ln \frac{D_n}{1 + \delta (D_n \sqrt{e} - 2)/m_{\mu}}\;\;,
                                                    \label{13}
\end{equation}
where $D_n = 1.54 A^{0.27}$, $A$ - mass number of the nucleus $(A = 56$ for
iron), $\delta = m^2_{\mu}E_{\gamma_\mu}/2E_1(E_1- E_{\gamma_\mu})$ --
minimal momentum transfer, and
$e=2.718$. The final numerical  results of eqs.~(\ref{10}) and (\ref{13}) are 
close to each other. In this paper we do not take into account the atomic
screening and inelastic form-factors, both nuclear and atomic (see
ref.~\cite{19}).
\begin{figure*}\flushleft
\begin{minipage}[htb]{0.48\linewidth}
\mbox{ \epsfig{file=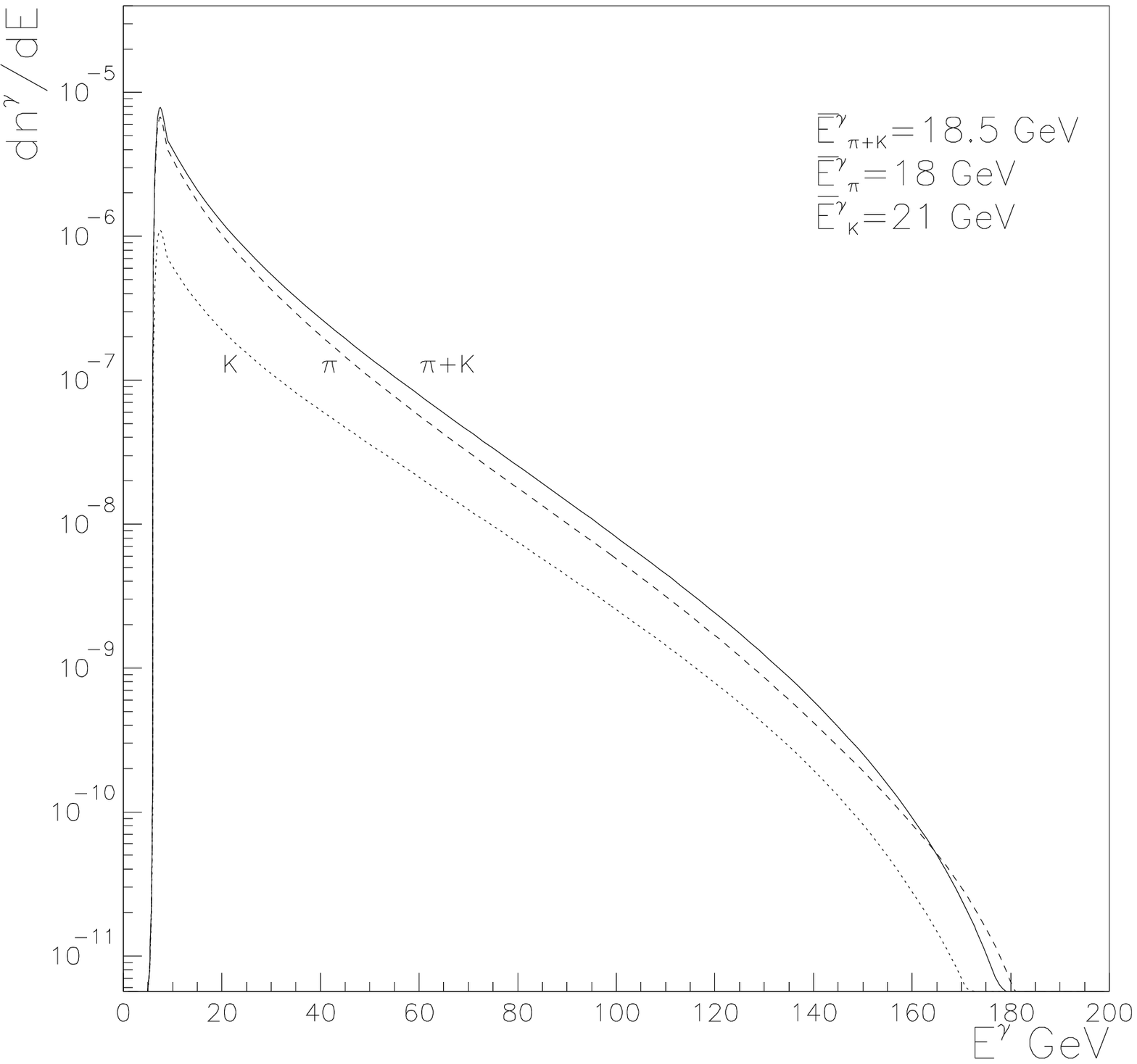,width=0.98\linewidth} }
\caption{
Spectra of muonic photons from external bremsstrahlung in the CERN 450 GeV
 WBB neutrino beam. The contribution from $\pi$-decay is shown as dashed line,
 from K-decays by dotted line.
 }
\label{gextsp}
\end{minipage}
\hspace{1mm}
\begin{minipage}[htb]{0.48\linewidth}
\mbox{ \epsfig{file=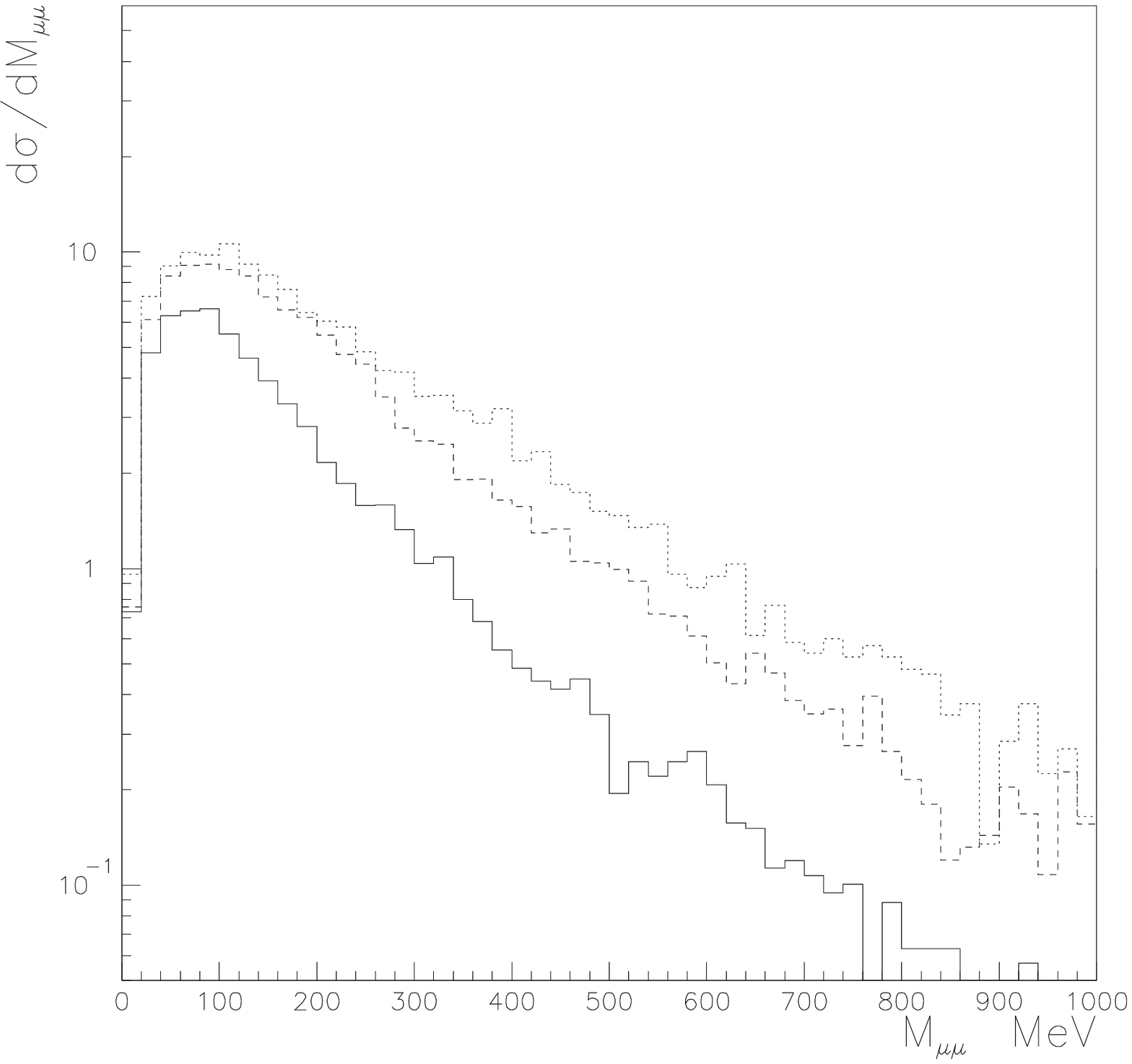,width=0.98\linewidth} }
\caption{
Invariant mass distributions (in units of pb/20 MeV) of muon pair
 from the conversion  
in the neutrino detector of muon gammas with fixed energies of 20 GeV (solid),
 50 GeV (dashed) and 100 GeV (dotted line). 
 }
\label{em45}
\end{minipage}
\end{figure*}

The total cross-section of the process (\ref{9}) convoluted with the spectrum
of muons emitted in pion and kaon decays
equals to 1760 pb for $\alpha_\mu=10^{-5}\alpha$. This
result was obtained  with the help of programs CompHEP and BASES. To calculate
the spectrum of muonic photons emitted by muons in the iron  shielding we used
the formula (\ref{10}) convoluted with the spectrum of muons.
The spectra of muonic photons, produced by the external bremsstrahlung
in the iron
shielding are presented in fig.~\ref{gextsp}.
The internal and external spectra are compared in fig.~\ref{gallsp}. 
First of all, one can see
that the yield of muonic photons produced by the internal bremsstrahlung is
higher for $m_{\nu_\mu}=10^{-10}$eV than the yield by the external
bremsstrahlung, supporting prospects to measure muonic neutrino mass if
$\gamma_\mu$ is discovered. The internal bremsstrahlung spectra are harder
than the external ones. Thus increasing the energy
of mesons (especially of kaons) should help in measuring neutrino mass. 
Moreover, the ratio of pion and kaon components is drastically 
different for muons (and hence for their
bremsstrahlung in the shielding) from that for the mesons themselves.  
Indeed, while initial
meson spectrum in the CERN beam has 80.4\% decayed pions,
 the muon
flux at the end of the decay tunnel has 96.5\%  muons emitted in the 
pion  decays. The relative
enhancement of pion component is caused by a strong suppression of the kaon
component, because the muons and muonic gammas from K-decays
have wider angular distribution and are partly missing the acceptance
 of the neutrino
detector  (see ~\cite{app:} for the  corresponding formulas).
However the contribution of energetic muons from kaon decays is enhanced by
rather high energy cut, 7 GeV, on muonic photons
 and the increase of the bremsstrahlung cross-section with energy.
Finally, 82.9\% of muonic photons and 80.3\% of muon pairs originate
from pions (this effect is discussed in
more details in ref. ~\cite{app:}).

In the external bremsstrahlung calculations we have neglected angular
 distribution of muonic photons
in the reaction (\ref{9}). Indeed, muons, moving at angles larger than
the angular size of the detector could, nevertheless, produce muonic photons
crossing the detector. Moreover, muons, moving in the directions crossing the
detector, could produce muonic photons going outside it. We assume that 
these two effects more or less 
compensate each other. The angular
distribution of muonic photons produced in the external  bremsstrahlung has a
forward peak with characteristic polar  angle $\theta \sim m_{\mu}/\bar E_1$,
where $\bar E_1$ is the muon average energy.  For $\bar{E}_1=31$ GeV
, more than 80\% of photons lie inside this $\theta$-cone 
(about $0.2^\circ$).
Thus, we assume that the effect under discussion should give
a correction factor not larger than 2 in  our final results.

\section{Photoproduction of muon pairs by $\gamma_\mu$'s. 
              \label{dimuon}}

The photoproduction of muons by ordinary photons has been analysed in
a number of papers.
The corresponding cross-section is connected with the bremsstrahlung
cross-section by crossing (see refs.~\cite{14,15}). One can use the
differential cross-section suggested in ref.~\cite{24}, substituting
$\alpha_{\mu}$ instead of $\alpha$, but keeping the factor $Z\alpha$
\begin{eqnarray}
    d\sigma(\gamma_{\mu}+Z \to \mu^++\mu^- +Z) 
    &=&
          \alpha_{\mu} \left(\frac{2Z\alpha}{m_{\mu}}\right)^2
          \left(1 - \frac{4x_+x_-}{3}\right)dx_+ \times \nonumber \\
    &\times& 
        \left[\ln \left(\frac{2}{3} \frac{m_{\mu}}{m_e}
          \frac{A_0}{Z^{2/3}}\right)
              - \ln \left(1 + \frac{A_0}{Z^{1/3}}
          \frac{\delta\sqrt{e}}{m_e}\right) \right].
                                                    \label{14}
\end{eqnarray}
Here $\alpha =1/137\;,\; Z$ is atomic number of the nucleus ($Z = 10$
for the `average nucleus' in the CHARM II detector), $\delta =
m^2_{\mu}/2E_{\gamma_\mu} x_+x_-$ -- minimal momentum transfer,
$E_{\gamma_\mu}$ -- the energy of the muonic photon, $x_{\pm} =
E_{\pm}/E_{\gamma_\mu}$, where $E_{\pm}$ -- energy of $\mu^{\pm}\;, \;
A_0 \simeq 190$ -- a constant, which determines the value of the
radiation logarithm.

\begin{figure*}\flushleft
\begin{minipage}[htb]{0.48\linewidth}
\mbox{ \epsfig{file=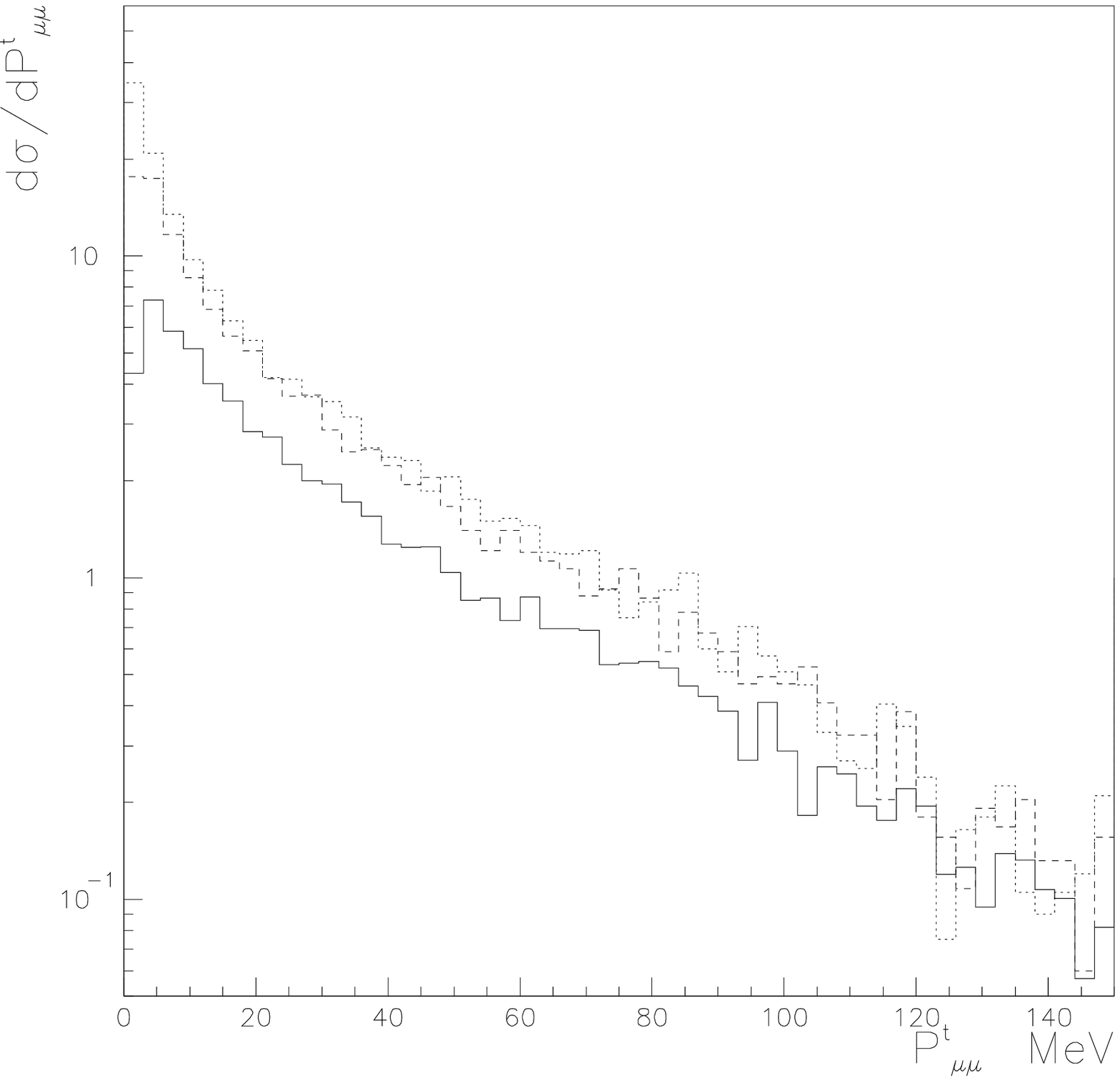,width=0.98\linewidth} }
\caption{
Transverse momentum distributions (in units of pb/3 MeV) of muon pair
 from the conversion  
in the neutrino detector of muon gammas with fixed energies of 20 GeV (solid),
 50 GeV (dashed) and 100 GeV (dotted line).
 }
\label{eptdis}
\end{minipage}
\hspace{1mm}
\begin{minipage}[htb]{0.48\linewidth}
\mbox{ \epsfig{file=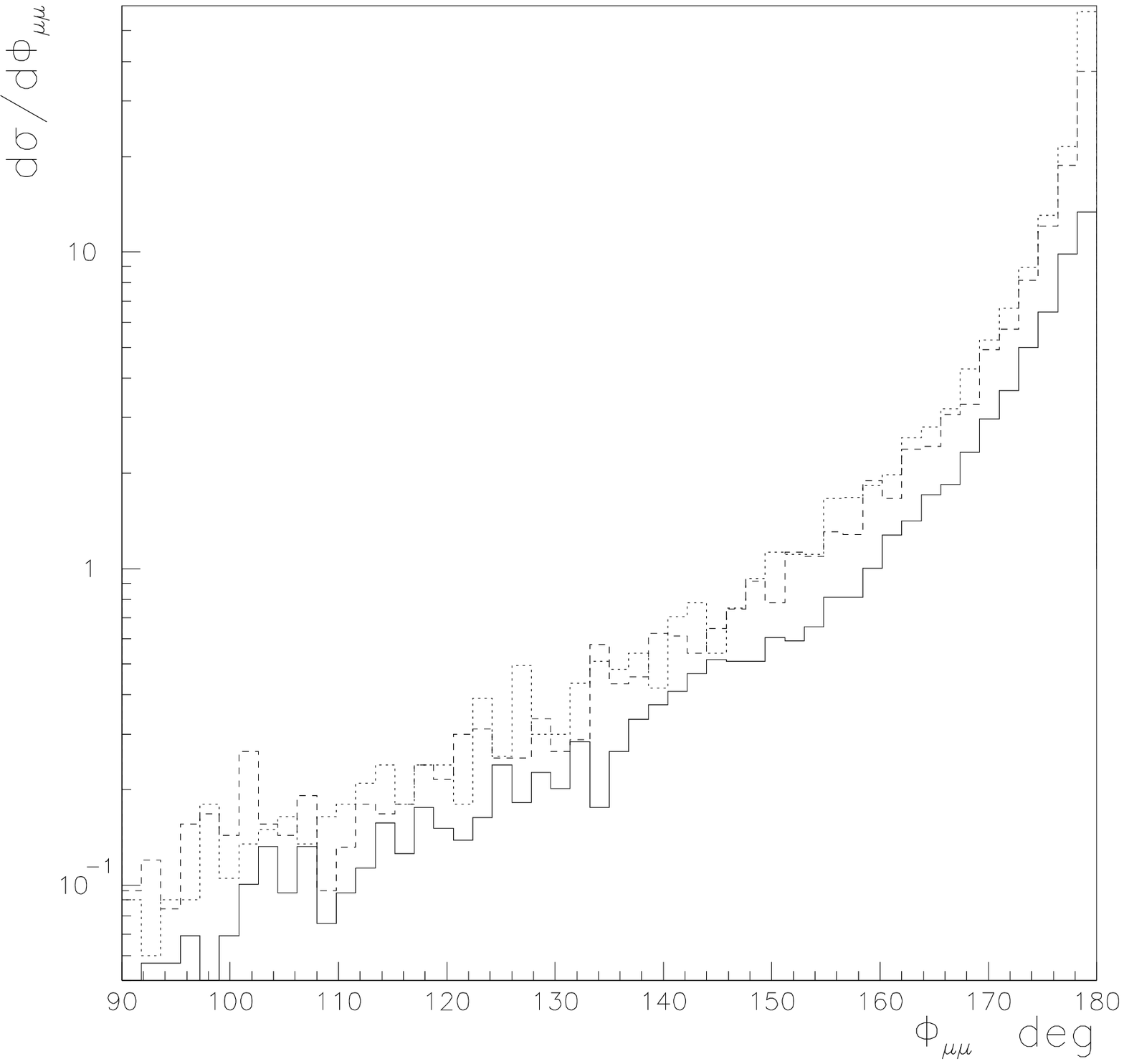,width=0.98\linewidth} }
\caption{
Azimuthal angle distributions (in units of pb/$1.8^{\circ}$) between two muons
 from the conversion  
in the neutrino detector of muon gammas with fixed energies of 20 GeV (solid),
 50 GeV (dashed) and 100 GeV (dotted line).
 }
\label{ephi}
\end{minipage}
\end{figure*}
With $E_{\gamma_\mu} \sim 20$ GeV and $0.1 <
x_{\pm} < 0.9$, the atomic screening may be neglected, so that the
logarithmic factor simplifies:
\begin{equation}
   \ln \frac{2}{3} \frac{m_{\mu}}{\delta Z^{1/3}\sqrt{e}}
     = \ln \frac{4}{3} \frac{E_{\gamma_\mu} x_+x_-}{m_{\mu} 
     Z^{1/3}}-\frac{1}{2}.
                                                     \label{15}
\end{equation}
Another, more recent parametrisation of the same cross 
section\footnote{S.R.Kelner, R.P.Kokoulin, private communication} is: 
\begin{equation} 
   d\sigma = \alpha_{\mu}\left(\frac{2Z\alpha}{m_{\mu}}\right)^2
             \left(1 - \frac{4x_+x_-}{3}\right) \
          \left[\ln\frac{ \frac{B}{Z^{1/3}}\frac{m_{\mu}}{m_e}}
          {1 + \frac{B}{Z^{1/3}}\frac{\delta}{m_e}\sqrt{e}} - 
          \ln \frac{D_n}{1+ \frac{\delta}{m_{\mu}} 
          (D_n \sqrt{e}-2)}\ \right] dx_+,
                                                    \label{16}
\end{equation}
where $B = 183,\; D_n = 1.54\cdot A^{0.27}$ and 
$A$ is a number of nucleons in a
nucleus ($A = 20$ for the `average nucleus' of the CHARM II detector).
Note that for not too small values of the $\delta$ parameter
($\delta/m_e\gg Z^{1/3}/B\sqrt{e})$ the square brackets in eq.~(\ref{16})
reminds the last factor of eq.~(\ref{10}) with substitution of the term 
$(-1.5)$ by eq.~(\ref{13}):
\begin{equation}
   \{\} \to \left[\ln\left(\frac{m_{\mu}}{\delta}\right) - \frac{1}{2} 
   - \ln\frac{D_n}{1 + \delta(D_n \sqrt{e}- 2)/m_{\mu}}\right].
                                                     \label{17}
\end{equation}
Note also that if we neglect  both atomic screening and nuclear
form-factor and integrate over $dx_+$, we reproduce an expression 
similar to the well known
result [ref.~\ref{15}, eq.~94.6]:
\begin{equation}
   \sigma = \alpha_{\mu}\frac{7}{9} 
            \left(\frac{2Z\alpha}{m_{\mu}}\right)^2 
            \left(\ln\frac{2E_{\gamma_\mu}}{m_\mu} - 
            \frac{109}{42}\right).
                                                    \label{18}
\end{equation}
Similarly to the case of the muon bremsstrahlung in the shielding the
nuclear elastic form-factor can be accounted for effectively by introducing
an additional term in brackets, $-1.2$. This value can be obtained from
the parametrisation (\ref{11}) for average nuclei with $Z=10$, `neon',
in the CHARM II detector.  Note that this term equals to $-1.5$ 
for Fe ($Z=26$), see (\ref{10}).

To get accurate numerical results for cross-section and to calculate
distributions for dimuon photoproduction we used the programs CompHEP
and BASES/SPRING .  In particular, the
total cross-section of dimuon photoproduction by an "average" muonic
photon, for $\alpha_\mu=10^{-5}\alpha$, is 41.5 pb in the case of
convolution with the normalized spectrum of muonic photons produced by
the external bremsstrahlung, and 35.6 pb in the case of the internal
bremsstrahlung (with $m_{\nu_{\mu}} = 10^{-10}$ eV).
  Note that in the case of internal bremsstrahlung the shapes of all
distributions depend only slightly on the
neutrino mass.  Of course this is caused by the similarity of muonic
photon spectra for different neutrino masses (see
fig.~\ref{gallsp}).  This similarity can be easily understood as
the main contribution comes from the logarithmic terms like that in eq. 
(\ref{7}).

Returning to the formula (\ref{18}) we note that its modified version,
with additional term $-1.2$ effectively accounting for elastic nuclear
form-factor, differs from the accurate calculation by CompHEP and BASES
by less than 1\% for $E_{\gamma_\mu}>50$ GeV, and this difference does
not exceed 10\% for lower energies within the energy range
$E_{\gamma_\mu}>7$ GeV discussed here.

Muons should have high enough energy to be detected; in our
calculations we took the detector energy threshold for muons as 4 GeV.
Keeping in mind this energy threshold we have applied energy cut 7
GeV  
on the energies of the original muons and neutrinos and on the energy of
muonic photons.  However for the final muons, produced in the detector by
muonic photons, we have applied cut $E_\mu>4$ GeV.  All values of
the photoproduction cross-section given above were obtained with these
cuts.

Taking into account the yields of muonic photons, normalized to one 
muon neutrino 
from two-body $\pi$/K mesons decays, calculated as discussed above 
(see~\cite{app:}),
\begin{center}
\begin{tabular}{ll} 
 $  3.1\cdot 10^{-7}$ & ext. brems.,\\
 $  1.5\cdot 10^{-7}$ & int. brems., $m_{\nu_\mu}=100$ KeV,\\
 $  10.4\cdot 10^{-7}$  & int. brems., $m_{\nu_\mu}=10^{-10}$ eV,
\end{tabular}
\end{center}
one can derive the the specific yields of muon pairs per unit neutrino flux:
\begin{center}
\begin{tabular}{ll} 
  $ 2.1\cdot 10^{-10}$ & ext. brems. events $\mbox{cm}^2$ /neutrino, \\  
  $ 0.9\cdot 10^{-10}$ & int. brems. events $\mbox{cm}^2$ /neutrino , 
                             $m_{\nu_\mu}=100$ KeV, \\
  $ 5.9\cdot 10^{-10}$  & int. brems. events $\mbox{cm}^2$ /neutrino,
                             $m_{\nu_\mu}=10^{-10}$ eV.
\end{tabular}
\end{center}
These specific yields of muon pairs  are not exactly proportional to the
specific yields of
muonic photons because of the difference of the mean energies and effective
cross-sections.
\noindent To obtain these numerical results we took the number of
average nuclei (with $Z=10$, `neon') in the CHARM II detector equal to
$0.16\cdot 10^{32}$ \cite{GBEAM}, and $\alpha_\mu=10^{-5}\alpha$.
Neutrino flux in CHARM II experiment is estimated \cite{trident} as
$8.6\cdot 10^{11}\,\nu/\mbox{cm}^2$ integrated over the whole energy
spectrum.
So we would expect 177 dimuon pairs in CHARM II exposure
from muonic photons produced by
external bremsstrahlung.  Then, in the case of internal
bremsstrahlung we should observe about 74 events, if the mass of
muonic neutrino is close to the experimental limit, $100$ KeV, and more
than  511 events, if the mass of muonic neutrino is very small, of order
$10^{-10}$ eV.
Invariant mass and transverse momentum of muon pairs, azimuthal angle
between two muons distributions are shown on
figs.~\ref{em45},~\ref{eptdis},~\ref{ephi} 
at some fixed energies of muonic photons:  20 GeV (solid lines),
 50 GeV (dashed lines) and 100 GeV (dotted lines).


The shapes of the transverse momentum and azimuthal angle distributions
scales with energy, while the invariant mass of two muons slightly
increases at high energy.

\begin{figure*}\flushleft
\begin{minipage}[htb]{0.48\linewidth}
\mbox{ \epsfig{file=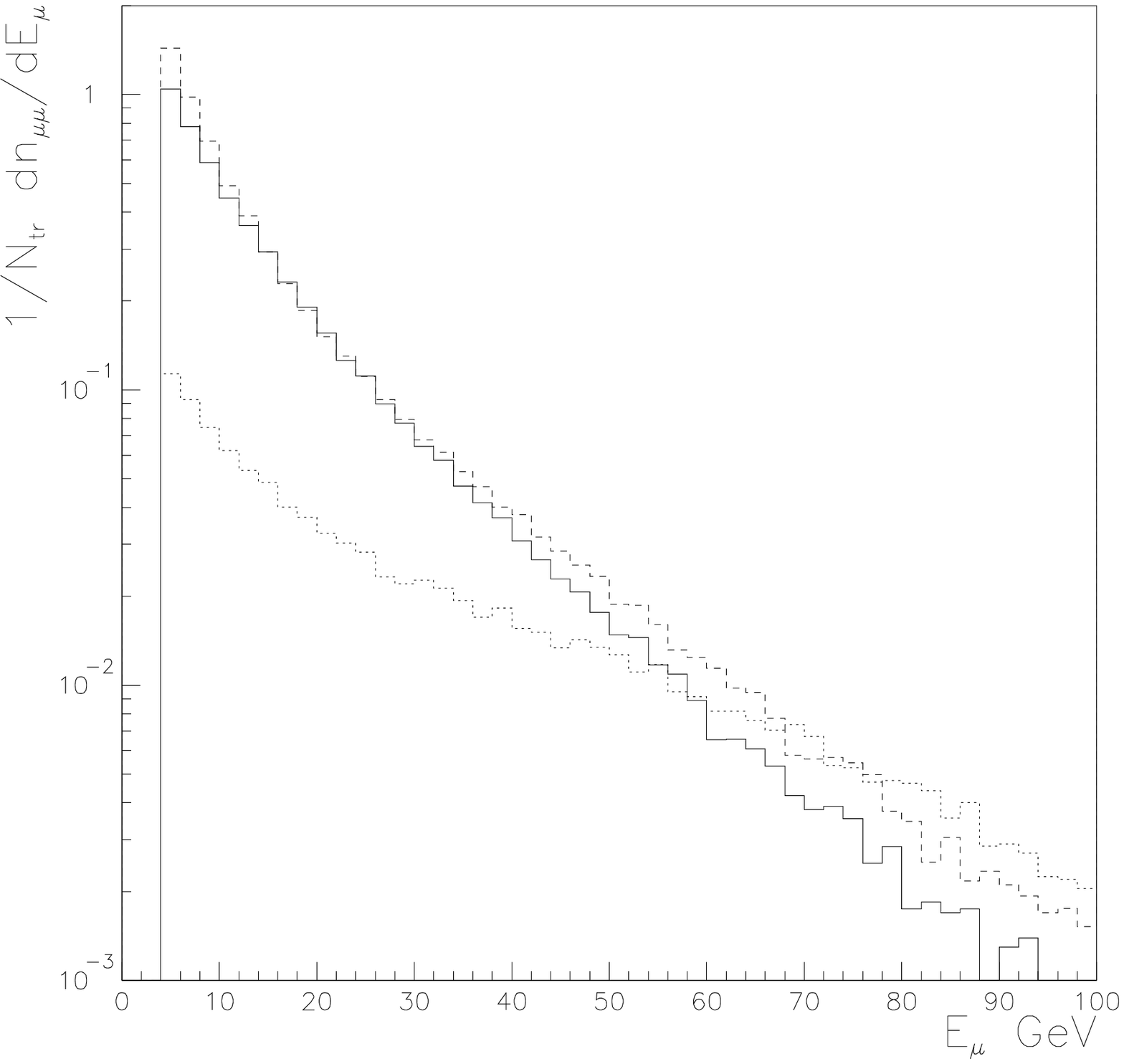,width=0.98\linewidth} }
\caption{
Energy spectra of muons
from the conversion of muonic photons and from tridents
in the neutrino detector at CERN
neutrino beam. Solid line corresponds to external brems, dashed line to internal
brems and dotted line to tridents. 
 }
\label{e4}
\end{minipage}
\hspace{1mm}
\begin{minipage}[htb]{0.48\linewidth}
\mbox{ \epsfig{file=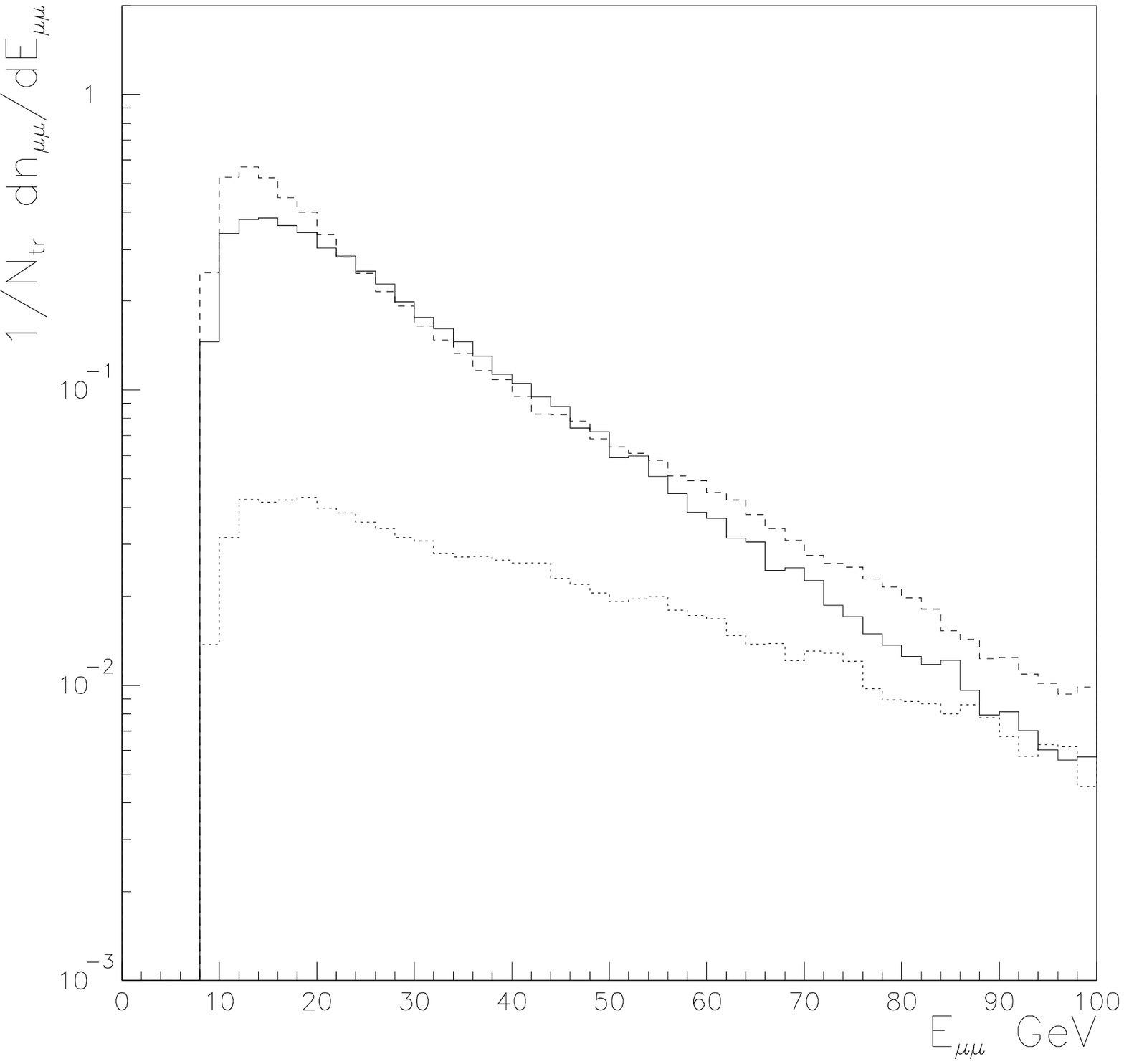,width=0.98\linewidth} }
\caption{
Energy spectra of muon pairs
from the conversion of muonic photons and from tridents
 in the neutrino detector at CERN
neutrino beam. Solid line corresponds to external brems, dashed line to internal
brems and dotted line to tridents.
 }
\label{e45}
\end{minipage}
\end{figure*}

The energy spectra of muons and muon pairs from tridents 
and from conversion of muonic photons
at CERN neutrino WBB are presented in figs.~\ref{e4} and \ref{e45}.

In these and following figures we represent external brems by solid lines,
internal brems with $m_{\nu_{\mu}}=10^{-10}$ eV by dashed lines and tridents
by dotted lines. All distributions are normalized to one trident event. 
One can note
that these spectra are harder in the case of the internal
bremsstrahlung than in the external one.

\begin{figure*}\flushleft
\begin{minipage}[htb]{0.48\linewidth}
\mbox{ \epsfig{file=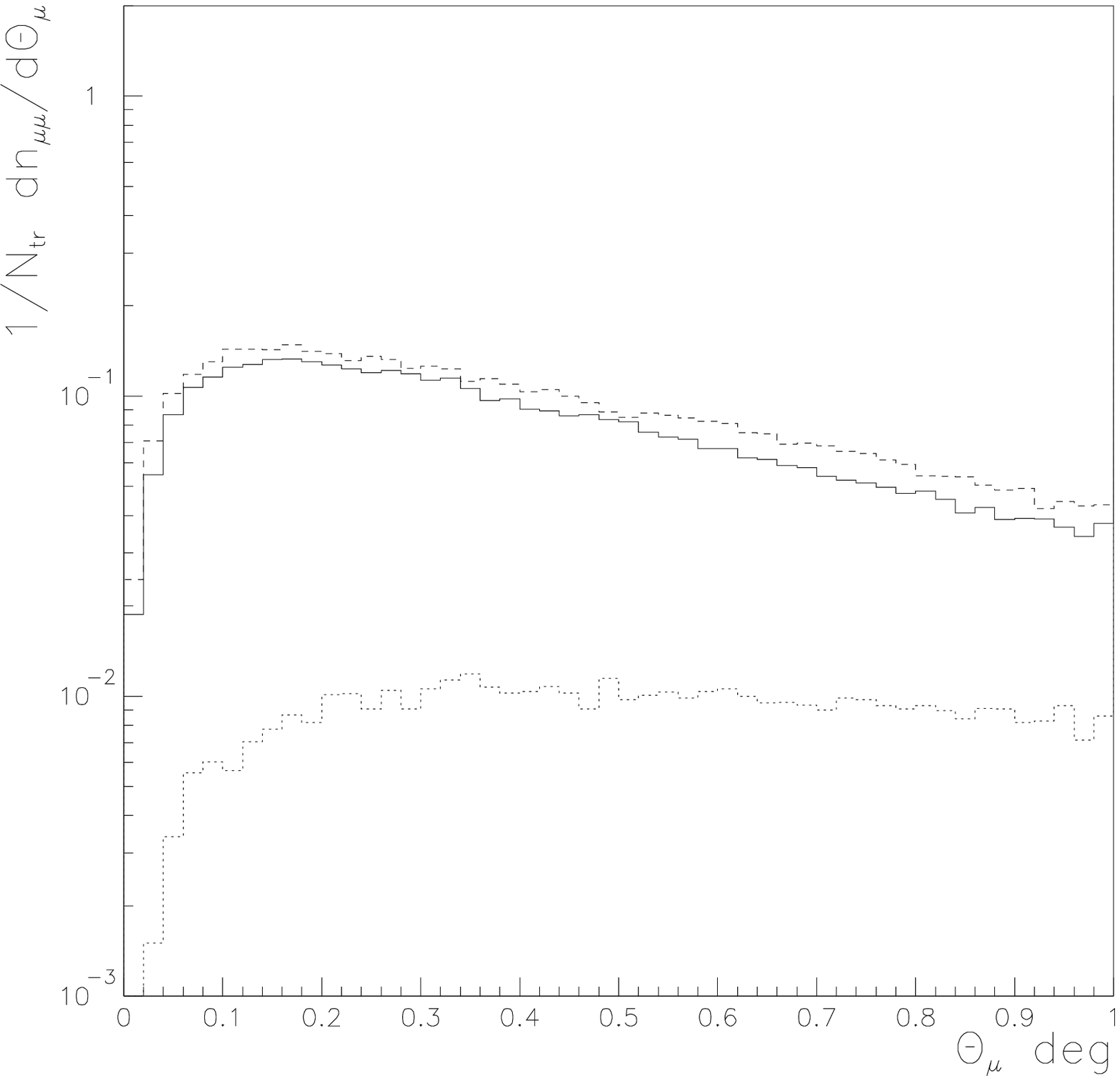,width=0.98\linewidth} }
\caption{
Angular distribution of muons
from the conversion of muonic photons and from tridents
 in the neutrino detector at CERN
neutrino beam. Solid line corresponds to external brems, dashed line to internal
brems and dotted line to tridents. 
 }
\label{th4}
\end{minipage}
\hspace{1mm}
\begin{minipage}[htb]{0.48\linewidth}
\mbox{ \epsfig{file=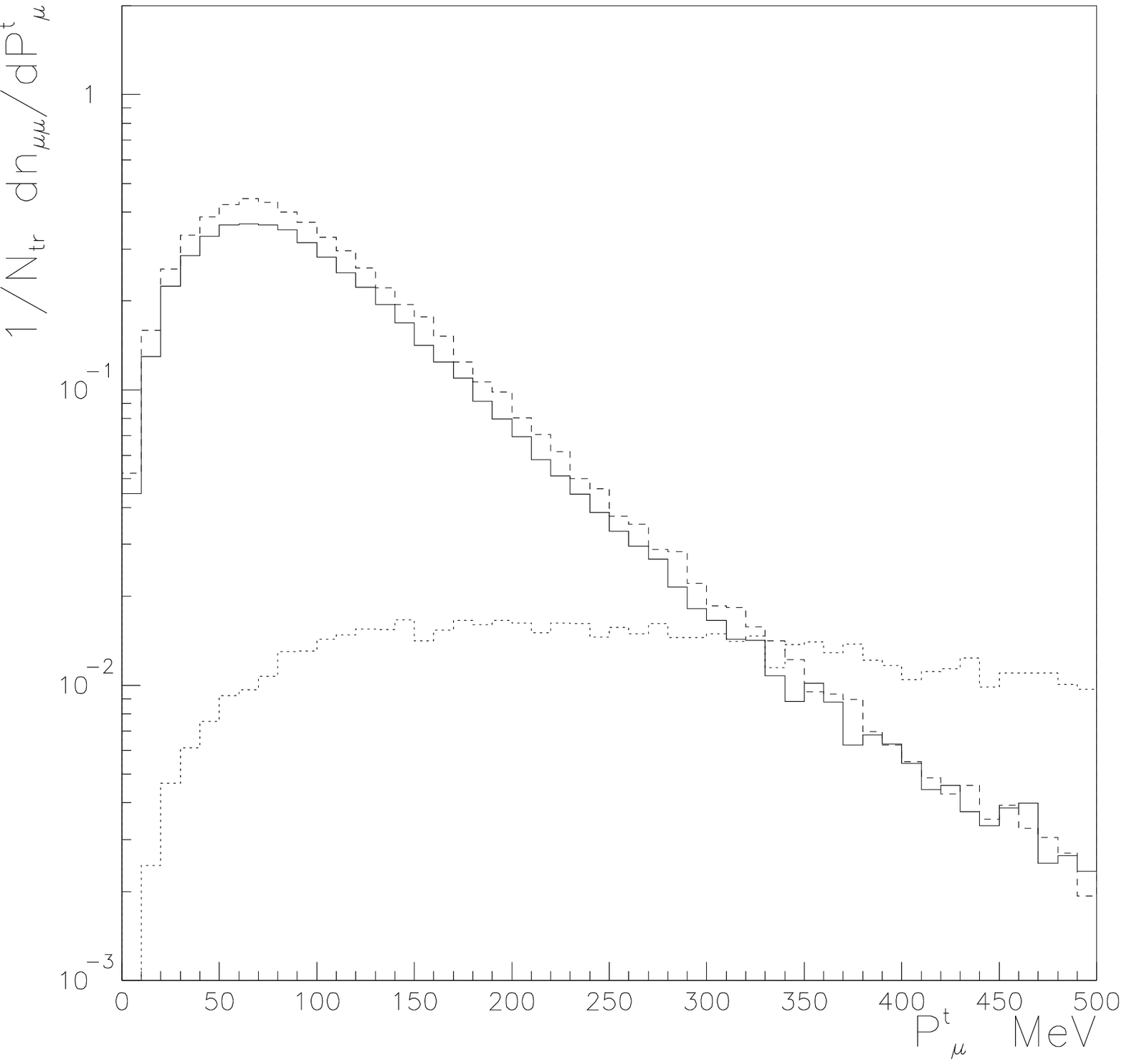,width=0.98\linewidth} }
\caption{
Transverse momentum of muons
from the conversion of muonic photons and from tridents
 in the neutrino detector at CERN
neutrino beam. Solid line corresponds to external brems, dashed line to internal
brems and dotted line to tridents.
 }
\label{q4}
\end{minipage}
\end{figure*}

One can see on fig.~\ref{th4} that muons from $\gamma_{\mu}$ conversions 
are produced at small
angles:  more than 80\% at angles less than $1^\circ$ with respect to
the direction of generic muonic photon.  We have found also that in
both cases (external and internal ones) muons are produced with rather
small $p_t$:  90\% of muons have transversal momentum less than 200
MeV (fig.~\ref{q4}).

\begin{figure*}\flushleft
\begin{minipage}[htb]{0.48\linewidth}
\mbox{ \epsfig{file=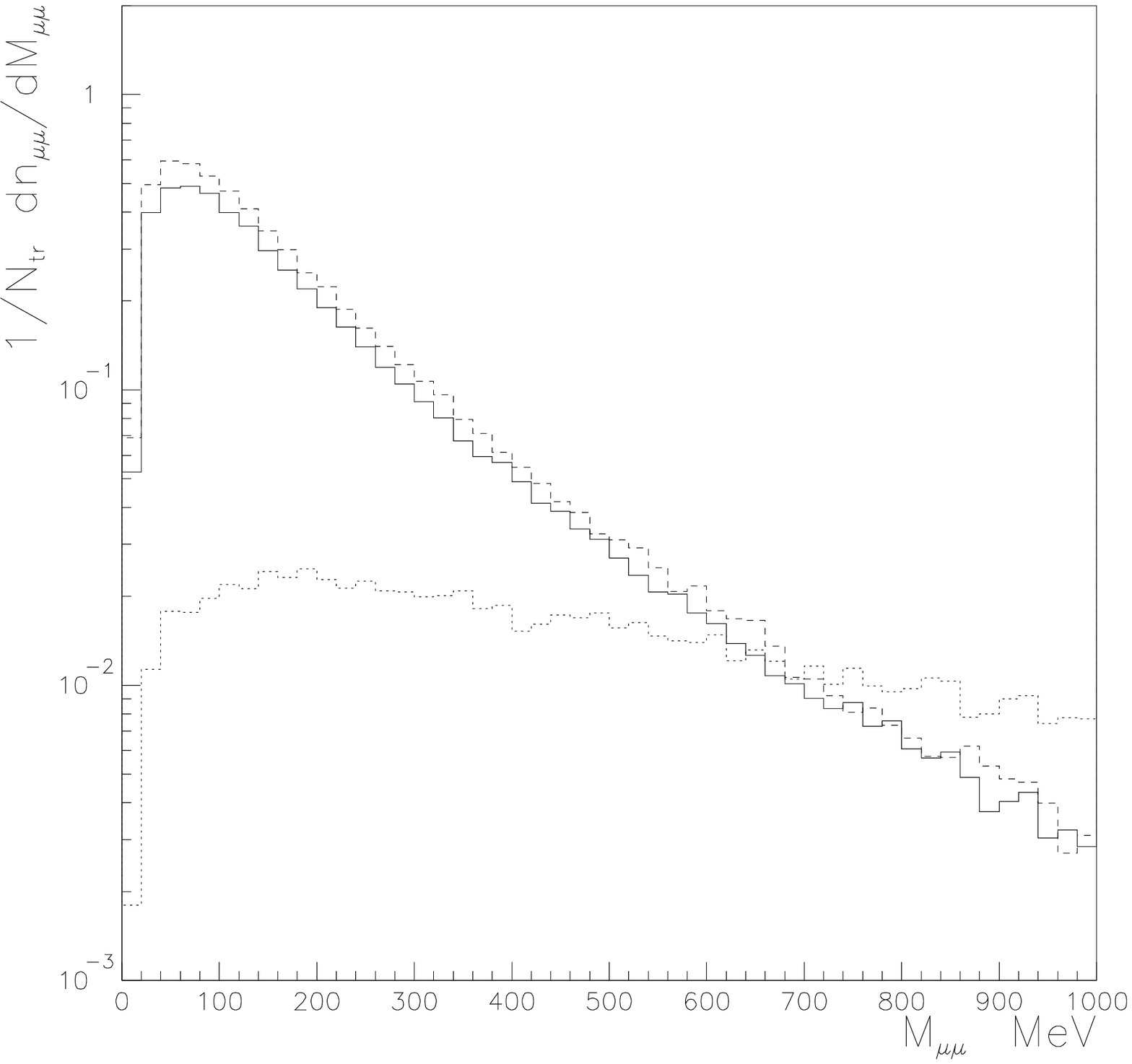,width=0.98\linewidth} }
\caption{
Invariant mass of muons
from the conversion of muonic photons and from tridents
 in the neutrino detector at CERN
neutrino beam. Solid line corresponds to external brems, dashed line to internal
brems and dotted line to tridents.
 }
\label{m45}
\end{minipage}
\hspace{1mm}
\begin{minipage}[htb]{0.48\linewidth}
\mbox{ \epsfig{file=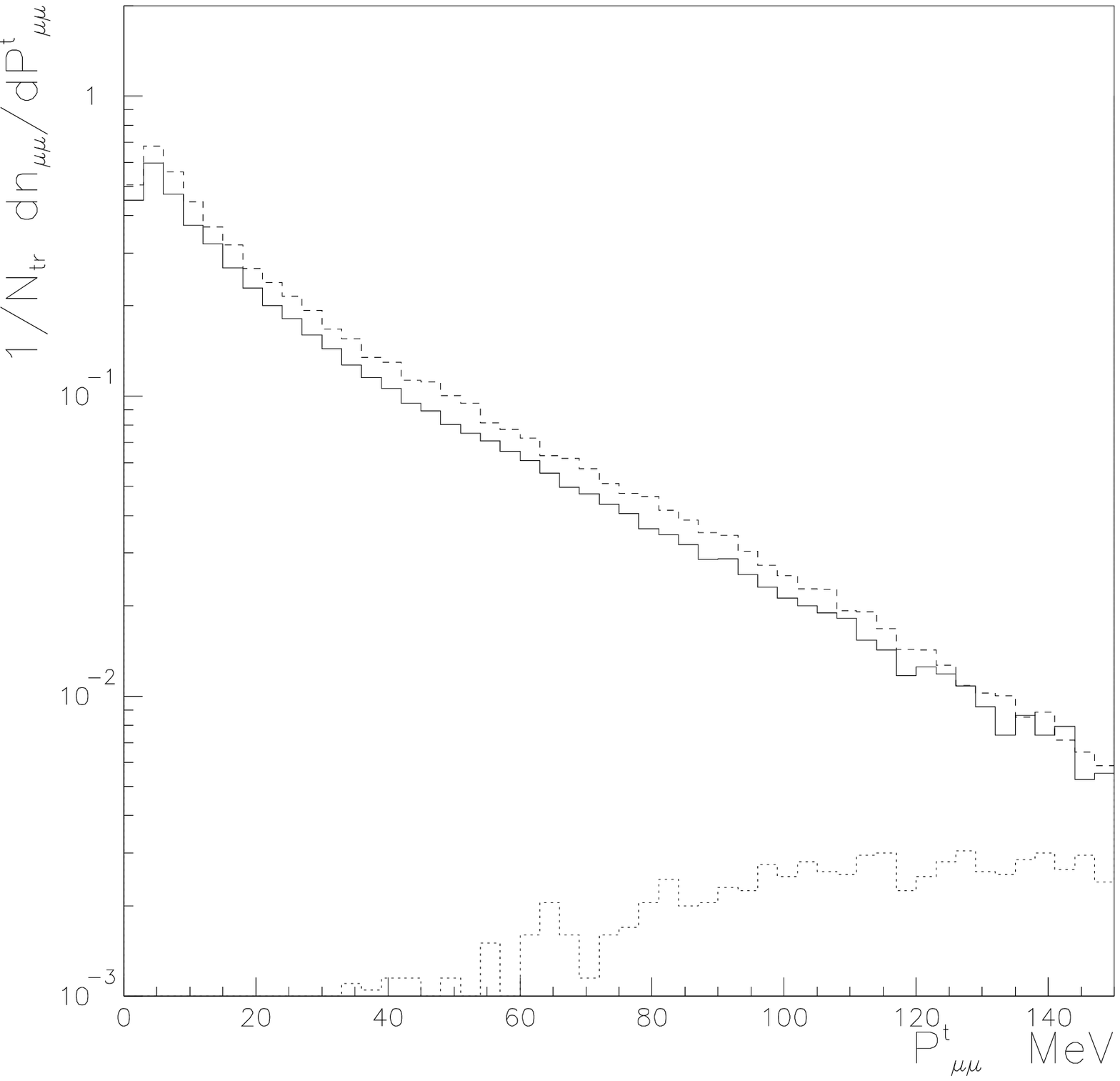,width=0.98\linewidth} }
\caption{
Transverse momentum of muon pair
from the conversion of muonic photons and from tridents
 in the neutrino detector at CERN
neutrino beam. Solid line corresponds to external brems, dashed line to internal
brems and dotted line to tridents.
 }
\label{q45}
\end{minipage}
\end{figure*}

\begin{figure*}\flushleft
\begin{minipage}[htb]{0.48\linewidth}
\mbox{ \epsfig{file=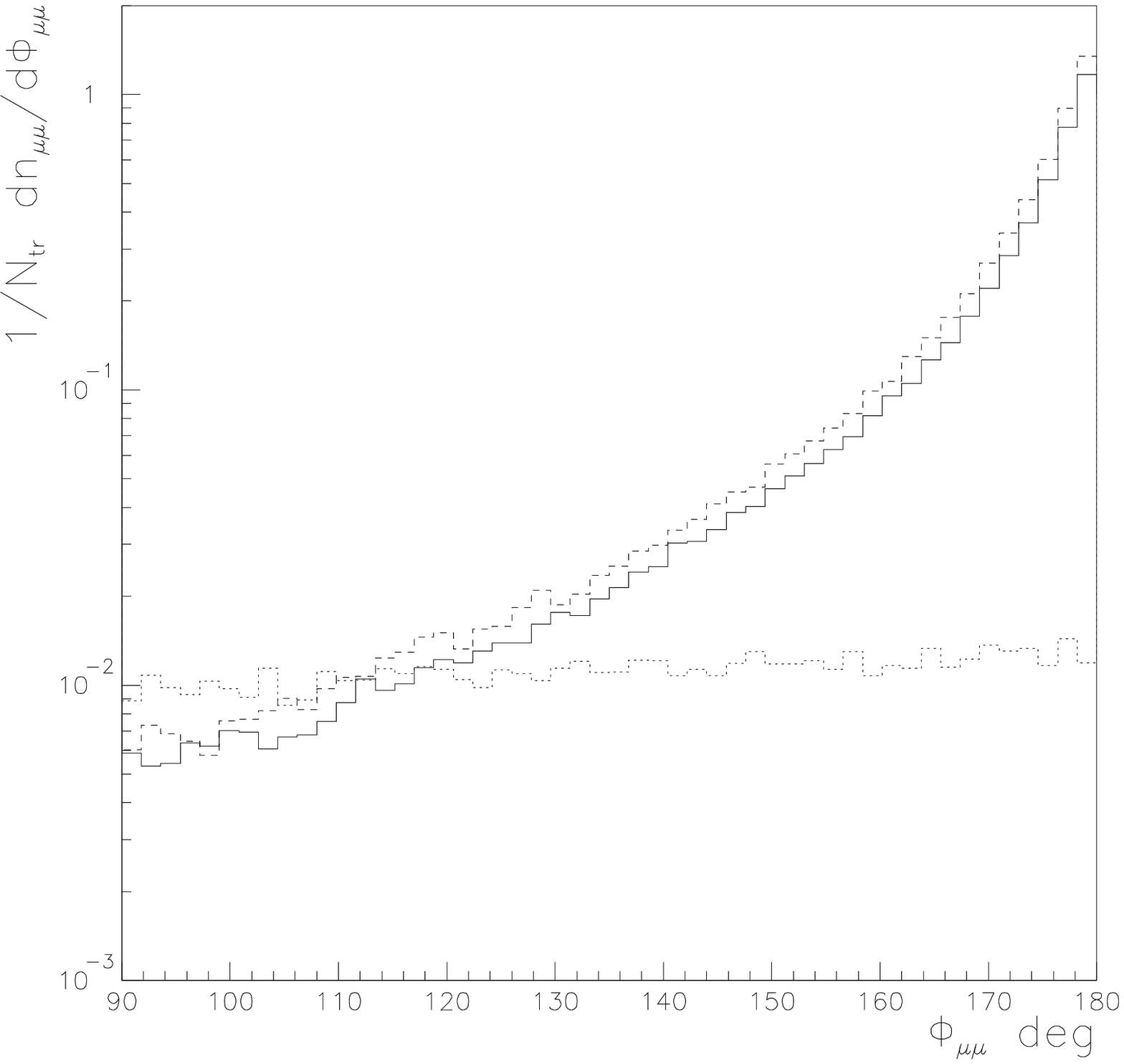,width=0.98\linewidth} }
\caption{
Azimuthal angle between muons
from the conversion of muonic photons and from tridents
 in the neutrino detector at CERN
neutrino beam. Solid line corresponds to external brems, dashed line to internal
brems and dotted line to tridents. 
 }
\label{phi45}
\end{minipage}
\hspace{1mm}
\begin{minipage}[htb]{0.48\linewidth}
\mbox{ \epsfig{file=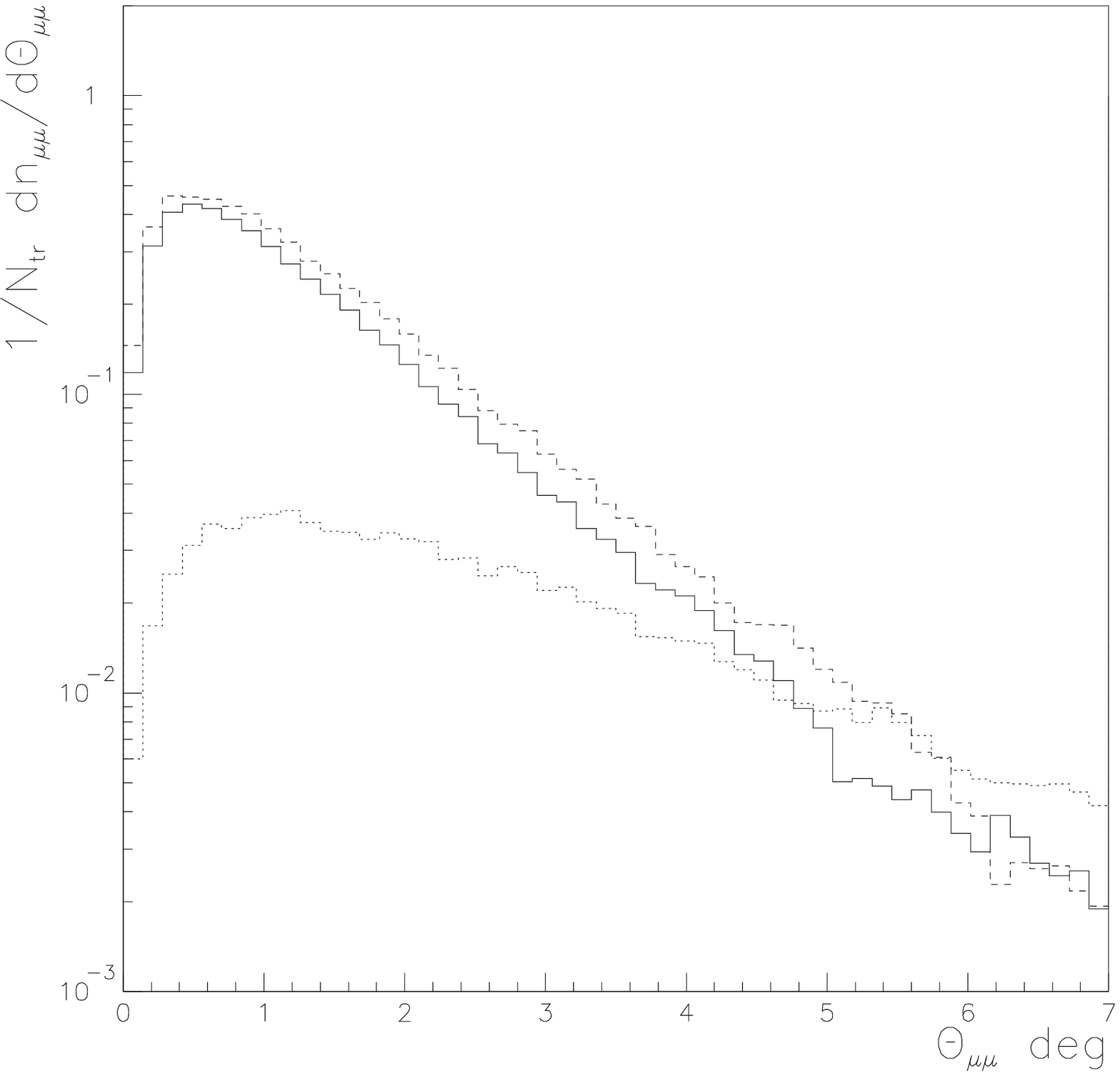,width=0.98\linewidth} }
\caption{
Angle between muons
from the conversion of muonic photons and from tridents
 in the neutrino detector at CERN
neutrino beam. Solid line corresponds to external brems, dashed line to internal
brems and dotted line to tridents. 
 }
\label{th45}
\end{minipage}
\end{figure*}

We also calculated distributions in different kinematical
variables to find most effective cuts to localize the effect
of muonic photons.
In figs.~\ref{m45},~\ref{q45},~\ref{phi45},~\ref{th45}
 we present the distributions in
the invariant mass of dimuons, in the transversal momentum of a muon
pair (the so-called {$p_t$-imbalance}), in the azimuth angle between
muons (it is an angle between the transversal components of $\mu^+$ and
$\mu^-$ momenta), and in the angle between total momenta of $\mu^+$ and
$\mu^-$.

\section{The trident background. \label{trident}}

As it has been mentioned in the introduction, the main background for
the search of muonic pairs produced by muonic photons is presented by
tridents, process (\ref{5}).  The cross-section of
trident production in a Coulomb field of a nucleus with charge $Z$, can
be easily estimated using Weizsaecker-Williams approximation (see e.g.
\cite{25}):
\begin{equation}
    \sigma(\nu_{\mu}+ Z \to \nu_{\mu} +\mu^++\mu^-+ Z) = 
          \frac{2(Z\alpha)^2 G^2}{9\pi^3} (g^2_L + g^2_R)\cdot 
          s_{max}\cdot \ln\frac{s_{max}}{4m^2_{\mu}}\;.
                                            \label{19}
\end{equation}
Here $G = 1.166\cdot 10^{-5}\,{\rm GeV}^{-2}$ is the four-fermion
coupling constant, $g_L=\frac{1}{2} + \sin^2 \theta_W$, \ $g_R = \sin^2
\theta_W$, \ $\sin^2 \theta_W = 0.23$, \ $s_{max} = 2 E_1 Q_{max}$, \
$Q_{max} = 1/a$, where $a$ is the radius of the nucleus, see
eq.~(\ref{11}); $E_1$ is the energy of the initial neutrino.  This
approximate formula gives the value smaller by $\sim 10$\% then the accurate
 calculation using CompHEP
and BASES within the discussed energy range.  The
cross-section convoluted with the normalized neutrino spectrum
 gives the value $\sigma
= 0.045\cdot 10^{-4}$ pb.  This numerical result has been obtained with
the nuclear form-factor (\ref{11}) and with energy cut $E_\mu>4$ GeV
applied on muons.
The number of
trident events is estimated by using the neutrino flux and
target mass given in Section~\ref{dimuon}
to be 50 events.

Energy spectra of muons, and muon pairs for tridents are presented 
by dotted lines in
figs.~\ref{e4} and \ref{e45}.  In
figs.~\ref{m45},~\ref{q45},~\ref{th45},~\ref{phi45}
 we present distributions (also by dotted lines) in
different variables for tridents --- in invariant mass of the muon
pair, in transversal momentum of the muon pair ($p_t$-disbalance), in
the angle between muon and antimuon, and in azimuth angle between muon
and antimuon.  One can see that trident events are distributed in these
variables broader than muon pairs in dimuon photoproduction.
  The characteristic values of kinematical cuts on these
variables are collected in table~\ref{tab:cuts}, together with the
percentage of the background trident events which survive these cuts.
It is obvious that the most promising is the
cut in the transversal momentum of muon pair,
$p_t$-disbalance. 
\begin{table}[htb]
\begin{center}
\begin{tabular}{|l|c|c||c|}
\hline 
 Observable  &  int.brems. & ext.brems.       & tridents \\
\hline
$M_{\mu\mu}$(MeV) 
  \raisebox{0ex}[3ex][3ex]{$<$}     & 500 & 500 & 39\%  \\ 
$p^{\mu\mu}_t$(MeV)  
  \raisebox{0ex}[3ex][3ex]{$<$}     & 72  & 73  & 2.12\% \\
$p^\mu_t$(MeV)    
  \raisebox{0ex}[3ex][3ex]{$<$}     & 200 & 200 & 24.9\%  \\
$\varphi_{\mu\mu}$(deg) 
  \raisebox{0ex}[3ex][3ex]{$>$}     & 142 & 142 & 25.6\%  \\
$\vartheta_{\mu\mu}$(deg) 
  \raisebox{0ex}[3ex][3ex]{$<$}     & 2.8 & 2.7 & 64.2\%   \\
$E_\mu$(GeV)      
  \raisebox{0ex}[3ex][3ex]{$<$}     & 28  & 28  & 65.8\%  \\
$E_{\mu\mu}$(GeV) 
  \raisebox{0ex}[3ex][3ex]{$<$}     & 56  & 52  & 69\%  \\
\hline
\end{tabular}
\end{center}
\caption{ 
The kinematical cuts, each of them preserves separately 90\% of muon
pairs produced by muonic photons in the detector (columns 2 and 3), and
the percentage of the trident background events within the cuts given
for external bremstrahlung case (column 4).  Here $M_{\mu\mu}$ is the
invariant mass of a muon pair; $p^{\mu}_t$ and $p^{\mu\mu}_t$ are
transversal momenta of a muon and a muon pair correspondingly,
$\vartheta_{\mu\mu}$ is the angle between $\mu^+$ and $\mu^-$, while
$\varphi_{\mu\mu}$ is the angle between their transversal momenta;
$E_\mu$ - energy of a muon ($\mu^+$ or $\mu^-$) and $E_{\mu\mu}$ is
energy of a muon pair.  For internal bremsstrahlung the numerical
results only slightly depend on neutrino mass within the discussed
range, at least up to $10^{-10}$ eV.
\label{tab:cuts}}
\end{table}
Only 2.12\% of trident events have
$p^{\mu\mu}_t< 70$ MeV.  By applying this cut one can hope to get the
suppression of trident background by a factor 50 while practically all
muon pairs from photoproduction will be kept.

The cuts on invariant mass of muon pair and on azimuth angle between
two muons could be also effective, each could suppress trident
background by a factor 3-4.  In addition, we found that
distribution in transversal momentum of muons 
is much broader for tridents than for dimuon photoproduction,
here the suppression
factor 4 could be also obtained for tridents. Of course special analysis
is necessary to find those combinations of these cuts which will give
maximal suppression factor for the trident background - the 
correlations are important.

\section{Monte-Carlo simulation of the production and detection of the muonic
        photons. \label{monte-carlo}}

In order to take into account the details of the geometry and kinematics of
the neutrino beam line, a special CERN Monte-Carlo program was used, which
directly simulates the decay of $\pi$/K with muonic photon and
the bremsstrahlung of muonic photons in the iron shielding
of the neutrino beam.
This program uses the Monte-Carlo $\pi$/K-decays produced by GBEAM program
\cite{GBEAM} with realistic geometry, magnetic focusing in the horn and reflector
and $\pi$/K lifetimes. So at the input every event has the coordinates of the
$\pi$/K decay point and the vector of $\pi$/K momentum. The difference with
the semi-analytical approach is that the radial and longitidinal spatial 
distributions of $\pi$/K in the decay tunnel, the angular distribution of 
$\pi$/K mesons and their correlations are automatically taken into account.
Both positive and negative $\pi$/K mesons are used for muonic photon flux
calculation in the neutrino Wide Band Beam (WBB),
 where positive mesons are focussed, 
while negative mesons are defocussed.
\begin{table}[htb]
\begin{center}
\begin{tabular}{|l|c|c|}
\hline 
 Observable  &  Semi-analytical & Monte-Carlo \\
\hline
 &  & \\
$\overline{E_{\gamma}^{int}}$ &  16.8 GeV & 17.9 GeV \\
$\overline{E_{\gamma}^{ext}}$ &  18.5 GeV & 22.8 GeV \\
$F_{\gamma}^{int}/F_{\nu}$ & $  10.4 \cdot 10^{-7}$ & $12.5 \cdot 10^{-7}$   \\
$F_{\gamma}^{ext}/F_{\nu}$ & $ 3.1 \cdot 10^{-7}$ & $2.3   \cdot 10^{-7}$  \\
$\sigma_{\gamma}^{int}$ &  35.6 pb & 39.9 pb  \\
$\sigma_{\gamma}^{ext}$ &  41.5 pb & 53.8 pb  \\
$N^{int}$ &  511 events & 688 events  \\
$N^{ext}$ &  177 events & 172 events  \\
$N^{tridents}_{total}$ &  140 events & 138 events  \\
$N^{tridents}_{4 \mbox{GeV} cut}$ &  49  events &  47 events  \\

\hline
\end{tabular}
\end{center}
\caption{ 
Results of semi-analytical and Monte-Carlo calculations
of the following quantities:
the average energies of muonic photons ($\overline{E_{\gamma}^{int}}$,
$\overline{E_{\gamma}^{ext}}$)
,
the ratios of muonic photon flux to neutrino flux for external
$F_{\gamma}^{int}/F_{\nu}$ and internal
$F_{\gamma}^{ext}/F_{\nu}$ bremsstrahlung,
the average muonic gamma interaction cross-sections for gammas from 
internal $\sigma_{\gamma}^{int}$ and external $\sigma_{\gamma}^{ext}$
bremsstrahlung, the numbers of events of muonic gamma conversions from internal
$N^{int}$ and external $N^{ext}$ bremstrahlung and the numbers of
 trident events without 
($N^{tridents}_{total}$) and with ($N^{tridents}_{4 \mbox{GeV} cut}$)
 the cut on muon momentum 
of 4 GeV. The fluxes, cross-sections and mean energies were calculated with
 the cut $E_{\gamma} > 7$ GeV and $E_{\mu} > 4$ GeV.
\label{tab:comp}}
\end{table}

 The internal bremsstrahlung of muonic gammas in $\pi$/K decays 
is generated according to 
the formulas used in the section \ref{intbrems}. The flux of the muonic
photons is calculated in the fiducial volume of CHARM-II detector with
square cross-section of 3.2x3.2 $\mbox{\rm m}^2$ located at 882 m from the proton
target.

The production of the muonic photons by external bremsstrahlung from muons,
stopping in the iron shielding was calculated according to the formula
(\ref{10}). For this study the shielding was appriximated by 80 blocks of iron,
 each 3m in length. The simplified tracking take into account the energy loss of
4.5 GeV per block and multiple scattering of muons in the iron. The muons were
 tracked down to 10 GeV. Muonic photons were generated with the angular distribution
(f.w.h.m. $0.12^{o}$) around the muon direction at each tracking step.
The effect of the multiple scattering reduce the flux of muonic photons in the
volume of the neutrino detector, because of the increase of the angular spread.

The conversion of muonic photons into muon pairs in the target of the
neutrino detector was simulated according to formula (\ref{16})
for the average CHARM II nuclei (A=20, Z=10). The energy cut of 4 GeV was
applied on both muons.

The cross-section and rates of the trident background were calculated by Monte-Carlo
program \cite{trident} with neutrino spectra predicted by GBEAM \cite{GBEAM} program.
Only coherent part of the cross-section was estimated for this study.

 The results of these calculations in the form of the mean
energies, cross-sections and rates are summarised in the table \ref{tab:comp}
for the neutrino mass $m_{\nu_{\mu}} = 10^{-10}$ eV and 
$\alpha_{\mu}/\alpha = 10^{-5}$.
 The numerical results of this Monte-Carlo
simulation are close to the semi-analytical calculations. The differences
are due to different approximations and assumptions on the geometry,
the spectra of mesons,
the effects of multiple scattering and the angular distributions.

\section{Conclusions. \label{conclusions}}

Let us summarize the main results of this paper. If  $\alpha_{\mu}\sim
10^{-5}\alpha$, as allowed by $(g-2)_{\mu}$ data, then in the high-energy
neutrino experiment of the type of CHARM  II one may observe $\sim 180$
 narrow
muon pairs produced by muonic  photons in the case of external bremsstrahlung.
In the case of internal bremsstrahlung there may observe about 70 events if
the mass of muonic neutrino is  $100$ KeV,
and $\sim 600$ events if the mass of muonic neutrino is very small, of
order  $10^{-10}$ eV. This should be compared with approximately 50 muon pairs
from trident events. Thus, muonic photons could be discovered in such
experiments  if  $\alpha_{\mu} \sim 10^{-5}\alpha$. Moreover, one may hope to
measure the mass of muonic neutrino if this mass is small enough and muonic
photon is discovered.

 By selecting pairs with $p_t<70$ MeV one may reduce the
number of background trident pairs by a factor of 50, to a few events.
One may hope to further suppress the trident background 
by selecting muon pairs with invariant mass less than 500
MeV,  or with  azimuth angle between muons larger than $140^\circ$, or with
transversal momentum of muon and antimuon smaller than 200 MeV.
Thus, if no events satisfying the cuts are observed, 
one can reach the sensitivity of $\alpha_\mu/\alpha<10^{-6}$.

\section*{Acknowledgments}

The authors are grateful to K.~Winter whose questions about the difference
between the muon pairs from muonic photons and tridents initiated this work.
L.O. is grateful to R.R.~Kokoulin, S.R.~Kelner, S.A.~Maltsev, A.Yu.~Morozov,
L.I.~Rozental, A.V.~Smilga, M.J.~Tannenbaum, V.R.~Zoller
 for interesting discussions, and to
RFBR for the grants 96-02-18010 and 96-15-96578. V.I. is grateful
to E.E.~Boos and A.E.~Pukhov
for useful discussions, and to the European association INTAS (contract
93-1180ext) and the Grant Center for Natural Sciences of State Committee for
Higher Education in Russia (grant 95-0-6.4-38).
A.R. is grateful to CHARM-II/CHORUS and CPPM colleagues
for interesting discussions and support.


\end{document}